\documentclass[12pt,preprint]{aastex}

\received{2001 November 6}
\begin{document}

\newcommand{\coto}{$^{12}$CO(2--1) \/}
\newcommand{\cooz}{$^{12}$CO(1--0) \/}
\newcommand{\thcooz}{$^{13}$CO(1--0) \/}
\newcommand{\thco}{$^{13}$CO \/}
\newcommand{\twco}{$^{12}$CO \/}
\newcommand{\vlsr}{$V_{LSR}$ \/}
\newcommand{\kmss}{km~s$^{-1}$ \/}
\newcommand{\kms}{km~s$^{-1}$}

\title{The Great PV Ceph Outflow: A Case Study in Outflow-Cloud Interaction}
\author{H\'ector G. Arce\footnote{Now at: California Institute of Technology,
Astronomy Department, MS 105-24, Pasadena, CA 91125. 
\newline New e-mail: harce@astro.caltech.edu} \/
\& Alyssa A. Goodman}
\affil{Harvard--Smithsonian Center for Astrophysics, 60 Garden St.,
Cambridge, MA 02138}
\email{harce@cfa.harvard.edu, agoodman@cfa.harvard.edu}

\begin{abstract}

We present a set of detailed molecular line maps of the
region associated with the
giant Herbig-Haro flow HH~315, from the young star PV Cephei, aimed
at studying the
outflow-cloud interaction.
Our study clearly shows that the HH~315 flow is effecting the
kinematics of its surrounding medium, and has been able to redistribute
considerable amounts of the surrounding medium-density
($\sim 10^3$~cm$^{-3}$) gas in its star-forming core
as well at parsec-scale distances from the source.
The single-dish observations include a map of the outflow
in the \coto line, with a
beam size of 27\arcsec, and more extended maps of the outflow
region in the \cooz and \thcooz lines, with 45\arcsec \/
and 47\arcsec \/ beam sizes, respectively. 
 A companion paper to this
one presents
higher-resolution (IRAM 30-m) observations, and discusses their
implications (Arce \& Goodman 2002). 
The giant molecular outflow HH~315 is a highly asymmetric
bipolar flow with a projected
linear extent of about 2~pc. Our results indicate that the two outflow
 lobes are each
interacting with the ambient medium in different ways. The southern
(redshifted) lobe, with a mass
of 1.8~M$_{\sun}$, interacts with a dense ambient medium, very close
to the young stellar outflow source, and its
kinetic energy is comparable to both the turbulent and
gravitational binding energy of its host cloud (of order $10^{44}$~erg). In
addition, we find evidence that
the southern lobe is responsible
for the creation of a cavity in the \thco emission.
In contrast, the northern (mainly blueshifted) outflow lobe, with a
total mass of 4.8~M$_{\sun}$,
extends farther from PV Ceph
and interacts with ambient gas much less dense
than the southern lobe. There is very little \thco emission north of
the outflow source, and the only
prominent \thco emission is a shell-like structure coincident with
the outer edge of the northern lobe, about 1.2~pc
northwest of PV Ceph.
It appears that the northern lobe of the HH~315 outflow has been able to
   ``push'' aside  a substantial fraction of the
gas in the area, piling it in a dense shell-like structure at its edges.
In addition, we find that about 50\% of the gas in the region of the
northern lobe has
been put in motion by the outflow, and that
the northern outflow lobe is responsible for a velocity gradient in
the ambient gas.

\end{abstract}

\keywords{ISM: jets and outflows --- ISM: Herbig-Haro Objects --- ISM:
individual (HH~315, PV Cephei) --- ISM: clouds --- stars: formation}

\section{Introduction}

As stars form inside molecular clouds, they gravitationally gather
gas from their surroundings,
   while simultaneously spurting out vast amounts of mass in
a  bipolar flow. The young stellar wind may reveal itself
in different ways. Among the
many manifestations of a nascent star's mass loss process are Herbig-Haro
(HH) objects and molecular outflows.
   An HH object is a nebulous knot, mainly seen at optical wavelengths, which
delineates the shock arising from the interaction of a high-velocity
flow of gas ejected by a young
stellar object (YSO) and the ambient medium. A chain of these HH
objects (or knots) is usually
referred as an HH flow. One or more shocks associated with an outflow
can accelerate entrained
gas to velocities greater than those of the quiescent cloud,
transferring momentum and energy into
the host molecular cloud, thereby producing a molecular outflow.

Since their discovery,  molecular outflows have been thought to
supply their host cloud with energy (e.g., Norman \& Silk 1980).
Yet, the (past) notion that outflows rarely extended to more than 1 pc
in length, and that they only effected a minor volume of the cloud,
made it difficult to believe that
stellar outflows could be important  to the energetics of their host
cloud. This picture is now changing.
Recent studies have shown that outflows from young stars  have an
enormous potential
to influence their parent cloud's structure and lifetime.
Optical and near-infrared observations have revealed that giant
---parsec-scale--- HH flows exist and
   that they are
common (Eisl\"offel \& Mundt 1997; Devine et al.~1997; Reipurth,
Bally, \& Devine 1997, hereafter
RBD; Mader et al.~1999; Eisl\"offel 2000; Stanke, McCaughrean, \&
Zinnecker 2000).
Giant HH flows have sizes about an order of magnitude larger than
the cloud cores from which they originate, and many are found to extend
to distances well outside the boundary of their parent dark cloud.
The colossal size of a giant HH flow enables it to
entrain molecular material at parsec-scale
distances from the source, and thus it may effect the kinematics
and density of a substantial volume of its parent molecular cloud.

In order to study the effects an outflow has on its environment's
kinematic and density structure,
it is imperative to observe a large area of the cloud gas
surrounding the outflow, in addition to the outflow itself.
The observations should preferably be of more than
one molecular line transition, probing a range of densities,
and at least one of the lines should
be relatively optically thin, in order to take line opacity into account.
Also, if one is to avoid source confusion, one should study regions
with a very low
density of young stellar objects.

The HH~315 flow,  from the young star PV Cephei (or PV Ceph)
is a perfect laboratory  to study outflow-cloud interaction 
as each outflow lobe interacts with drastically different environments.
In addition, there is little chance for source confusion in the area 
near HH~315 since there 
are no known outflow-powering
YSOs, other than PV Ceph, in a 30\arcmin \/ radius  centered on PV
Ceph.

The giant HH~315 flow was discovered independently
by G\'omez, Kenyon \& Whitney (1997, hereafter GKW) and RBD.
The presumed source of the HH 315 flow is PV Cephei,  a
variable Herbig Ae/Be pre-main sequence star located at
the northeastern edge of the L1158 and L1155
group of dark clouds, which is believed to be at a
distance of 500~pc (Cohen et al.~1981).
A brief literature review of the observations of PV Ceph is given by
GKW. The HH~315 flow (see Figure~1) consists of
six major condensations (which are named knot A, B, C, D, E, F), and
each major condensation
(or knot) is made of two or more individual small HH knots
(identified F1, F2, F3, etc.).
In addition, a group of knots,  which were
originally discovered by Neckel et al.~(1987),
are seen 20\arcsec \/ to 55\arcsec \/ north of PV Ceph, and are
identified as HH~215.
RBD state that HH~215 and HH~315 are all part of a single  2.6~pc-long (in
projection) HH flow which, all  together, are commonly referred to as
the HH~315 flow.
The northern lobe of the HH~315 flow is made of HH~215, and knots A, B, and C.
The southern lobe is made of knots D, E, and F (see Figure~1).
The spectra of several knots (Devine 1997) show that the northern lobe is
blueshifted and the southern lobe is redshifted.

RBD affirm that the distances from the
source to knot A, B, and C are within 10\% of the distance from the
source to knots
D, E, and F, respectively (see Figure~1).
They argue that the coincidence in the distance from the source
to the knots, and the S-shaped point symmetry of the two lobes around
the source (see below)
are evidence that the six knots come from three major mass ejection
eruptions. In this picture the oldest eruption is responsible for the
C-F knot pair, followed by the
eruption responsible for the B-E pair, which is followed by the
eruption responsible for the
A-D pair. The most recent eruption is responsible for the HH~215
knots, and a counter-knot
to HH~215 is not observed south of PV Ceph
presumably because is heavily obscured by the dense gas near PV Ceph (RBD).
Assuming a  tangential velocity of 200~\kmss for each knot, RBD
estimate that the major
eruptions in PV Ceph have occurred every 2000 yr, starting with the
C-F eruption about 6500 yr
ago.

A very peculiar aspect of HH~315 is that the knots trace a roughly
sinusoidal path, or
S-shaped point symmetry around PV Ceph.
RBD and GKW argue that this particular flow morphology strongly
suggests that the ejection
axis of PV Ceph is wandering or precessing (i.e., changes over time).
GKW modeled the HH~315
flow with a simple precessing jet model. In their model the jet axis
precesses on a cone with full
opening angle, $\theta$, inclined to the plane of the sky at an angle
$i$, and the jet is assumed to have
a constant velocity and a constant precession velocity.
GKW state that assuming a jet velocity of 200~\kms, with $i \sim 10\arcdeg$,
$\theta \sim 45\arcdeg$, and a precession  period of about 8300 yr
their models can produce a jet consistent with the HH~315 flow
morphology.  It is also possible
to explain the knot position angles and symmetries with a model where
PV Ceph is moving at a
substantial velocity ($\sim10$ \kms) and pairs of knots are emitted
at varying angles as the source
travels through space (see Goodman \& Arce 2002).

The molecular outflow associated with the HH~315  flow was first
observed by Levreault (1984).
The low resolution \cooz map of  Levreault (1984) shows that the
HH~315 molecular outflow
is asymmetric. The redshifted lobe has a circular morphology with a
diameter of $\sim 0.3$~pc, and the
blueshifted lobe has an elongated morphology which extends up to $\sim 1.2$~pc
northwest from PV Ceph.
Here we present new, more sensitive, and higher spatial resolution
$^{12}$CO(2--1), $^{12}$CO(1--0), and \thcooz observations
of the HH~315 molecular outflow, its surroundings, and a large
fraction of its parent dark cloud. Our analysis focuses on the
effects of the giant HH~315 outflow on its host cloud.

In Section 2  we give a description of our observations,
and in Section 3 we describe our results.
In Section 4 we discuss the mass and energetics of the HH~315 flow and the
effects that the HH~315 flow has on its parent cloud.
We devote Section 5 to our conclusions.
   A study of the kinematics of the HH~315 outflow and the entrainment
mechanism is discussed in Arce \& Goodman (2002, hereafter Paper II) ,
using even higher angular and velocity resolution (IRAM~30~m telescope)
molecular line data of the gas immediately
surrounding PV Ceph (the HH~215 region) and the HH~315B and  HH~315C 
optical knots.

\section{Observations}

\subsection{$^{12}$CO(2--1)  Data}

The \coto line data were obtained using
the on-the-fly mapping technique at the
National Radio Astronomy Observatory (NRAO) 12~m
telescope on Kitt Peak, Arizona, in December 1998.
At the observed frequency of 230~GHz, the telescope's
half-power beam width, main beam efficiency, and aperture efficiency
are 27\arcsec \/, 0.32 and 0.44, respectively.
The spectrometer used was a filter bank with
250 kHz resolution, with two independent sections of 128 channels each.
The filter bank was put in parallel configuration, in which each
of the two sections received independent signals with a different
polarization. The parallel configuration was chosen so that ultimately
the two polarizations could be averaged to produce better signal-to-noise
spectra. At a frequency of 230~GHz, the resultant velocity resolution for this
setup is 0.65~km~s$^{-1}$.

The on-the-fly (OTF) mapping technique was used to observe an
area of about 215\sq\arcmin \/ surrounding
the HH 315 flow. The OTF technique
allowed the extended area to be mapped in a more efficient way than the
conventional point-by-point mapping. The telescope in OTF mode
moved across the source at a constant speed of 30\arcsec~s$^{-1}$,
while a spectrum was acquired every 0.1 seconds.
In order to map the desired area more
efficiently, the total area was mapped by combining three
overlapping regions of different sizes and orientation.
The largest region (hereafter central region) has a size of 11\arcmin \/
$\times$ 13\arcmin,
oriented at a position angle of -26\arcdeg \/  centered around the position
of PV Ceph, at R.A.~$20^h45^m23^s.3$, decl.~67\arcdeg46\arcmin36\arcsec \/ (B1950).
   The second region (hereafter SE region) is a rectangle of
about  10.5\arcmin \/ $\times$ 6.5\arcmin, and it is centered southeast
of PV Ceph at R.A.~$20^h45^m58^s$, decl.~67\arcdeg40\arcmin15\arcsec \/ (B1950).
The third region (hereafter NW region) is a rectangle, of about
10.5\arcmin \/ $\times$ 6.5\arcmin, and it is centered northwest of PV Ceph at
R.A. $20^h44^m52^s$, decl. 67\arcdeg53\arcmin00\arcsec \/ (B1950).
Each region was scanned in both
the directions perpendicular and parallel to its long axis.
The separation, in the direction
perpendicular to the scanning direction, between subsequent rows was
7.2\arcsec \/
for the central region and 8\arcsec \/ for the SE and NW regions.
The telescope was pointed to an OFF position after
every other row, where it would observe the OFF position for 10 sec and then
vane calibrate for 5 sec. We used a different OFF position, depending on the
region observed in order to reduce the time needed to move from the ``ON''
area to the OFF position. The central region and the NW region used the same
OFF position, located at R.A. $20^h45^m26^s$, Dec. 67\arcdeg56\arcmin22\arcsec
\/ (B1950) and for the SE region we use a different OFF, located at
R.A.  $20^h45^m26^s$, Dec. 67\arcdeg56\arcmin22\arcsec \/ (B1950).
Deep observations at the OFF positions show that there is very faint
emission at these positions
in the velocity range between 0.5 and 3.0~km~s$^{-1}$,
with $0.1\lesssim T_{A}^{*} \lesssim 0.3$~K,
and no emission greater than 0.1~K at other velocities.
Most of the outflow emission is at velocities less than and greater than
0.5 and 3.0~km~s$^{-1}$, respectively.
Therefore, the very small amount of emission at velocities
between 0.5 and 3.0~km~s$^{-1}$ at the OFF positions does not affect
our outflow data.
The different regions were observed several times to improve the
signal-to-noise in the spectra.
The system temperature was measured to be in the range
between 350 and 800~K.
Our goal was to maintain a constant noise level through the whole map, so
regions that were mapped with a higher system temperature were given
comparatively more integration time.

The raw OTF data were reduced using  various AIPS OTF tasks.
A first-order baseline was fitted and subtracted to each spectrum,
and the two polarizations were averaged. The different mapped regions were
combined and averaged. The map was then convolved onto a grid with
14\arcsec \/ pixels (Nyquist sampled). The resultant RMS noise
in each 0.65~km~s$^{-1}$ channel was 0.1~K, for most (non border) spectra.
The intensity scale of the spectral
data presented in this chapter is in units of $T_{A}^{*}$ (Kutner \&
Ulich 1981) unless
otherwise stated.

\subsection{$^{12}$CO(1--0), $^{13}$CO(1--0),
and C$^{18}$O(1--0) Data}

In order to study the effect of the HH~315 outflow on a larger scale
and to obtain
a better estimate of the outflow mass, we observed
the \cooz line in a region
17.6\arcmin \/ by 29.3\arcmin \/ and the
\thco line in  a region 11.7\arcmin \/ by 23.5\arcmin, both along a position
angle of -30\arcdeg, surrounding PV Ceph.
The data were obtained using the SEQUOIA 16 element
focal plane array receiver of the Five College Radio Astronomy Observatory
(FCRAO) 14~m telescope. The observations were done over the course of three
different observing rounds, which took place in April 1999, December 1999,
and February 2000.
The backend used for observing both lines was the Focal Plane Array
Autocorrelator Spectrometer, with a channel spacing of 78~kHz
(0.21~km~s$^{-1}$) for $^{12}$CO(1--0), at a frequency of 115~GHz,
and a channel spacing of
20 kHz (0.05~km~s$^{-1}$) for $^{13}$CO(1--0), at a frequency of 110~GHz.
The \cooz line was observed in position switching mode, with a single OFF
position located at R.A. $20^h46^m50^s$, decl.
68\arcdeg01\arcmin36\arcsec \/ (B1950).
Observations at the OFF position reveal no significant ($T_{A}^{*}
\lesssim 0.2$~K)
emission.
The \thco line was observed in frequency switching mode.
The telescope half-power beam widths for the \cooz and \thcooz lines
are 45\arcsec \/ and 47\arcsec, and the main beam efficiencies are
0.45 and 0.54, respectively. Both lines  were observed on a
22\arcsec \/ grid (Nyquist sampled), and
with an integration time of 100~sec for each position.
The system temperature of our $^{13}$CO(1--0) observations was about
250~K, and for
our $^{12}$CO(1--0) observations it
ranged between 500 and 900~K.

Observations of the C$^{18}$O(1--0) emission were also made using the same
telescope and backend configuration as the $^{13}$CO(1--0) observations.
We made one 5.5\arcmin \/ $\times$ 5.5\arcmin \/ FCRAO-SEQUOIA
footprint centered at the position of PV Ceph.
The telescope half-power beam width and main beam efficiency for
C$^{18}$O(1--0), at a frequency of 109~GHz, are 47\arcsec \/ and 0.54,
respectively. Each position was observed for 210~sec, with a system
temperature of about 280~K.

The \cooz and \thcooz
data were reduced with the CLASS and MIRIAD packages. For the purpose
of obtaining the
molecular gas mass (see \S 3.4), the original
spectra from both lines were smoothed and resampled to a
channel spacing of 0.22~km~s$^{-1}$. The spectra were spatially linearly
interpolated and a data cube of 11.5\arcsec \/
pixels was produced for each line. The data cube was then smoothed
by convolving it with a Gaussian with a FWHM of 46\arcsec.
 The resultant RMS noise of the
spectra in the \cooz and \thco smoothed maps
is 0.14~K and 0.04~K for each 0.22~km~s$^{-1}$ channel, respectively.
Interpolating the \cooz and the \thco data to this common position-velocity
grid allowed us to use them in concert when calculating the mass estimates
presented in \S 3.4.

\section{Results}

\subsection{$^{12}$CO(2--1) Maps}

\subsubsection{Integrated Intensity Map}

In Figure~1 we show the integrated velocity map of the molecular outflow
associated with the HH~315 flow, created by integrating over wide
blue- and red-shifted velocity ranges.
The integrated velocity map is overlaid on an optical image of the
region (from RBD).
The large extent of the flow is clearly seen.
The blueshifted
lobe (northwest of PV Ceph) is about 0.6~pc wide and 0.8~pc long, and
the redshifted lobe is
about 0.3~pc wide and 0.5~pc long.\footnote{
The full flow, measured as 
the distance
between the outermost HH knots, is  2.6 pc long (see Figure 1).}   Another
striking aspect of this molecular outflow, besides its large extent, 
is the difference
in  size and morphology of the
two outflow lobes.
The blue lobe shows a bow shock-like appearance, with an axis of P.A.
$\sim -26\arcdeg$, and  with
its northern edge coincident with the optical knot HH~315C. Most of
the blueshifted
emission resides in a region near the optical knot HH~315C and
HH~315B, and does not
extend all the way to source. In contrast, most of the red lobe's
emission is concentrated in
a collimated north-south structure that extends from PV Ceph south
about 0.3~pc. The redshifted
lobe also includes a more extended, wider region of low intensity
emission which has
its southern edge near the position of the optical knot HH~315D.
In addition to the morphological differences, the lobes also have different
maximum outflow velocities.
Our large scale map, shows \coto blueshifted lobe velocities up to
$\sim 16.0$~\kmss away
from the central velocity of the ambient cloud. On the other hand,
the \coto red lobe  shows
maximum  outflow velocities of only up to $\sim 6.5$~\kms.

Levreault (1984) suggested that the asymmetry in the lobes'
morphology and velocity distribution
is due to the underlying density distribution in the PV Ceph region.
Our \thcooz map of the region (see \S 3.2) shows that there is indeed a 
drastic difference in the ambient density between north and south 
of PV Ceph. The outflow-cloud interaction should be a momentum-conserving
interaction, thus the difference in ambient density may partly explain the
asymmetry between the molecular outflow lobes' morphology and kinematics.
 
It is interesting to note that although the {\it molecular} outflow shows a
high degree of asymmetry between its two lobes, 
the lobes of the HH flow are highly symmetric about the source 
(see Figure 1). 
A clump of gas propagating
through the interstellar medium (e.g., an HH knot) will
decelerate by drag forces and the deceleration is expected to be 
 proportional to the density of the ambient medium
(De Young \& Axford 1967; Cant\'o et al.~1998; Cabrit \& Raga 2000). 
Thus, one would expect the southern HH knots of HH~315 to decelerate  
 more than the northern HH knots, resulting in an asymmetric HH flow. 
We believe that HH~315 is able to keep its high degree of 
symmetry because the southern HH knots only travel a short distance 
inside the high density regions of the  molecular cloud. The southern 
HH knots must travel through dense gas once they are 
ejected near the source. However, our  high-resolution data
  (see Paper II for details) indicate that the HH knots 
are quickly out of the molecular cloud
and coasting through low-density atomic gas
by the time they reach the position of HH~315E (at a distance
of 0.9~pc from PV Ceph).

\subsubsection{Channel Maps}

In Figure~2
we show  12 velocity-binned (``channel") maps of the \coto data. It is clear
that most of the
\coto emission comes from the (0.65~\kms-wide) velocity channels
centered at 0.38, 1.03, 1.68, 2.33,
and 2.98~\kms. The central
velocity of the ambient gas associated  with PV Ceph is about
2.5~\kmss (see Cohen et al. 1981, and \S 3.3), 
thus most of the ambient cloud emission surrounding
PV Ceph comes from the channels centered at 2.33 and 2.98~\kms.
The emission at these two velocity channels concentrates around (and south of)
the location of PV Ceph, but there is also a local peak in the
emission about 9\arcmin \/
northwest of the source.

It should be noted that in our \coto map we also detect emission from another
   cloud which is unrelated
to the cloud associated with the young star PV Ceph. This cloud
(hereafter cloud X)
   contributes a major fraction
of the total emission seen at  $v_{chan}=1.03$~\kms,
and is also detected at $v_{chan}=0.38$ and
$v_{chan}=-0.28$~\kmss (see Figure~2).
Cloud X has a filamentary structure with  a northwest-southeast long axis,
most clearly seen in  $v_{chan}=0.38$~\kms, and may also be seen,
less clearly at
$v_{chan}=1.03$ and -0.28~\kms.

The blueshifted lobe of the PV Ceph molecular outflow is easily discernible
at  $v_{chan}=0.38$~\kms, where the structure of the \coto gas about
9\arcmin \/ northwest
of the source, has a bow shock-like appearance.
This bow-shock-like
structure is coincident with the position of the HH~315C optical knot
(see also Figure~1),
and can also be clearly seen at the velocity channels centered at
-0.28,  -0.93,
and -1.58~\kmss, and less clearly at $v_{chan}=-2.23$~\kms.
At $v_{chan}=-0.28$~\kms, this structure is about 0.5~pc wide 
and 0.75~pc long.

This blueshifted CO bow morphology arises from the interaction between 
a bow shock from an underlying jet (i.e., the HH~315C optical knot) and 
the ambient gas (see Paper II for details).
The average axis of the HH~315 flow is believed to be
close to the plane of the sky (GKW). Thus, it is reasonable to assume 
that the redshifted emission at a distance of $\sim$ 9\arcmin \/
northwest of the source,
near the position of the HH~315C knot, at
$v_{chan}=2.98$ and $v_{chan}=3.63$~\kmss (see Figure~2),
comes from the gas accelerated away from us, on
the ``back side'' of the HH~315C bow shock (see Paper II).

At higher blueshifted velocity channels ($v_{chan}=-0.93, -1.58,
-2.23$~\kms) some of the
bow shock-like structure can be seen near HH~315C, in addition to
emission at the positions of the optical knots HH~315A and HH~315B.
The blueshifted gas emission near HH~315A is very faint, and
dispersed. On the other hand,
the \coto associated with the HH~315B blueshifted optical knot is
relatively strong and
has a defined knot-like structure, with a peak \coto
emission coincident with the location of the peak in the HH~315B
optical emission (see Figures 4.1 and
4.2).

The redshifted velocity channel maps show a very different morphology than the
blueshifted channels. At $v_{chan} \geq 3.63$~\kmss most of the \coto
emission is limited to a
small region close to PV Ceph.  In $v_{chan}=3.63$~\kms, in addition
   there is  also some faint \coto
emission near the position of the redshifted optical knot HH~315D,
   and some faint emission near HH~315C.
At  $v_{chan} \geq 4.28$ all of the emission is concentrated very
close to the source,
mostly in a north-south structure.

The \cooz velocity maps show basically the same structure as the
\coto maps. Our \cooz data set is
not as sensitive as our \coto data set, thus for illustrative purposes we
only show the \coto maps.

\subsection{$^{13}$CO(1--0) Maps}
Most of the \thcooz emission in this region of the sky is optically
thin, and thus is a good
probe of the structure of the medium density ($n \sim 10^3$~cm$^{-3}$) gas.
The \thcooz opacity may be estimated with C$^{18}$O(1--0) observations of the
same region (e.g., Heyer et al.~1987).
Our 5.5\arcmin \/ by 5.5\arcmin \/ map of C$^{18}$O(1--0) centered on PV Ceph
   shows that the C$^{18}$O emission
   is concentrated around the young star,  at velocities between 2.1
and 3.2~\kms,
and that there is no detectable emission at other velocities nor
at distances greater than 2\arcmin \/ from PV Ceph.
The maximum emission in our \thco map (Figure~3) is near PV Ceph, and 
hence it is fair
to assume that all other local peaks in the \thco emission far from PV
Ceph are less optically thick
than the emission near PV Ceph.
Thus, we assume that
the \thco emission in the area we studied is only mildly optically thick
very near PV Ceph and other local \thco maxima
   (and only at ambient cloud velocities) and  it is optically thin
everywhere else.
On the other hand, the \twco emission is optically thick almost everywhere
(see \S 3.4), so the \thco emission is a much better
probe of the cloud structure.

In Figure~3 we show six \thco maps, integrated over six different
velocity ranges. 
  It should be noted that the velocity ranges shown in each panel 
of Figure~3 are
not of equal width.  The velocity ranges were chosen to 
group different channels with similar \thco emission structure.

The first map is integrated over the velocity range of
$0.50<v<1.50$~\kmss (Figure~3{\it a}). At these  velocities
   most of the emission is concentrated west
and northwest of PV Ceph.   
An interesting feature seen in Figure~3{\it a} is the shell-like structure
about 9\arcmin
\/ northwest of PV Ceph. This structure encloses the bow shock-like
blueshifted \coto lobe of the
HH~315  molecular outflow (see Figures~1 and 2). It appears that the
blueshifted lobe of the
HH~315 flow has ``pushed'' aside the $^{13}$CO, piling the gas
in a shell-like structure at the edges of the blue lobe (see \S 4.2.2
for a discussion of this).
Note that Cloud X can also be seen in \thco emission at these velocities.

Figures~3{\it b} and 3{\it c} show the $^{13}$CO integrated emission over
the ranges $1.50<v<1.84$~\kmss and $1.84<v<2.17$~\kms, respectively.
The \thco emission in Figure~3{\it b} is concentrated near the western edge of
our map. The \thco emission in 
Figure~3{\it c} is also mainly concentrated near
the western edge of the map, but, it is more spread out (towards the
east) than the emission
seen in Figure~3{\it b}. It is at the velocities of Figure~3{\it c}
($1.84<v<2.17$~\kms) where we start to detect
\thco  associated with the cloud that harbors PV Ceph (see below).
The \thco emission over  the velocity ranges $2.17<v<2.61$~\kmss and
$2.61<v<3.05$~\kmss (Figures~3{\it d}
 and 3{\it e}, respectively) show the main \thco
structure of the cloud associated with the PV Ceph young star.
The cloud's northern edge is just north of PV Ceph, and the cloud extends
mainly  towards the south-southwest of the young star.
The \thco structure at these velocities is very similar
to the \coto structures seen at $v_{chan}=2.33$ and 2.98~\kmss (see Figure~2).

In Figure~3{\it f} we show the \thco integrated intensity map over the velocity
range $3.05<v<3.48$~\kms, which reveals the most redshifted  \thco
emission of our map.
All of the
\thco emission  is concentrated at the position of PV Ceph. This (unresolved)
redshifted \thco emission most probably comes from
the medium density gas  very near PV Ceph that has been accelerated by the
HH~315 flow, or could also be due to the motion of PV Ceph (see
Goodman \& Arce 2002).

\subsection{Dissecting the PV Ceph cloud}

This subsection is intended to explain certain definitions we will use
in the discussion that follows.  Table~1 summarizes the definitions.
We use the term ``PV Ceph cloud'' to refer to the cloud associated
with the formation
of the young star PV Ceph.
We define the area and velocity limits of the
PV Ceph cloud based on the \thco density and velocity structure.
The central velocity of the PV Ceph cloud is 2.5~\kmss
(see below, and Cohen et al.~1981),
and the slowest (detected) redshifted molecular outflow velocity is
0.7~\kmss from the central velocity (see \S 3.4). Thus, we assume that
0.7~\kmss ``blue-ward''
and 0.7~\kmss  ``red-ward'' of the central velocity, that is  $1.8 < v
< 3.2$~\kms, is the PV Ceph ambient cloud emission velocity range. 
It can be seen,
from the \thco velocity channels,
that it is at these velocities ($1.8 < v < 3.2$~\kms) where there is
\thco emission surrounding PV Ceph,
which does not come from cloud X.

In Figure~4 we plot the
\thco position-velocity diagram of the mapped region,
constructed by summing all spectra along an axis with
a position angle of -26\arcdeg \/ over the whole width of the \thco map.
In Figure 4, the lower dashed box delimits the ``PV Ceph Cloud," and 
the upper box
defines the ``North Cloud'' (see Table 1).
   The box  defining the PV Ceph cloud
avoids the
high-velocity \thco feature at velocities greater than 3.2~\kms. The
box also avoids the
   \thco emission at velocities lower than
1.8~\kms.
This emission at $v<1.8$~\kmss is due to the \thco clump
south-southwest of PV Ceph (see Figure~3{\it b}) which
 is not part of the cloud associated with the formation of PV Ceph.
The small velocity gradient south of PV Ceph  is
due to the kinematical distribution of the large scale \thco emission 
in our map. 

The integrated intensity map
of  the \thco emission  integrated over the PV Ceph ambient cloud velocities 
($1.8 < v < 3.2$~\kmss) is shown in Figure~5.
   The eastern and northern edges of the cloud
are easily defined by the drop in \thco emission (see Figure 5 and Table 1).
The western and southern edges are not so clear,
but it seems that R.A.  $20^h44^m32^s$ and decl. 
67\arcdeg36\arcmin30\arcsec \/ (B1950)
are  reasonable choices for the eastern and southern edges, respectively..
   Figure~3 shows that  most, if not all,
of the \thco emission at the defined  velocities of the PV Ceph cloud
($1.8 < v < 3.2$~\kms) lies east of
$20^h44^m32^s$ and north of 67\arcdeg36\arcmin30\arcsec \/ (B1950).

In Figure 6 we show the area that we later  use to calculate the mass of each
molecular outflow lobe (see \S 3.4).
We also show in Figure~6 (in dashed squares)
the area used to estimate the velocity
dependence of the \cooz line opacity
for each outflow
lobe region, as discussed in \S 3.4. These smaller areas were
chosen to have CO emission
representative of the larger  (outflow mass) areas, with as little 
contamination from
unwanted cloud (and velocity) structures (e.g., cloud X) as possible. The
``ambient" velocity for the Southern lobe, $v_{amb, south}$, 
is obtained from a
two-Gaussian fit to the average
\cooz spectrum shown in Figure~7{\it a}.  
The resultant value, for the
narrow velocity component  is $v_{amb, south}=2.5$~\kms.

It is evident that the  molecular gas north of PV Ceph
exhibits a large-scale velocity gradient.
The overall gradient is in the sense that  the  \twco and \thco emission (with
cloud-like  morphology, presumably non-outflow gas) is shifted to the 
blue towards the
north-northwest of the map (see Figures 2, 3, and 4). The central 
velocity of the northern
lobe region (or ``Northern Cloud") is
$v_{amb, north}=1.5$~\kms, obtained using the same procedure 
described above for
$v_{amb, south}$. We define the ``Northern
Cloud'' to be in the velocity range $0.8 < v  < 2.2$~\kmss, and to 
have an area shown by
the box in Figure~6 (see also Table 1 and Figure~4).
Later, we argue that the sense of the overall velocity gradient 
north of PV Ceph is not an accident, and
that most of the mass in the northern cloud  has been accelerated to 
blueshifted
velocities by the HH~315 flow (see \S 4.2.2).

\subsection{Outflow Mass}

\subsubsection{Procedure}

To obtain the outflow mass we use the method described in \S
3.3.2 of Arce \& Goodman (2001b, hereafter AG).
 This method, which is based on the method employed 
by Bally et al.~(1999) and Yu et al.~(1999), 
 uses the \cooz to \thcooz ratio
to estimate the opacity in the \cooz line, as a function of velocity.
In most cases the \cooz line is optically thick, and its opacity varies with
velocity.  Typically, the highest-velocity gas in the line wings is 
optically thin,
and the optical depth increases as velocities approach the line 
center.  Using an
optically thick line, without properly correcting for its  velocity-dependent 
opacity,
will result in an underestimation of the the outflow mass, momentum and kinetic
energy.

Here we give a brief description of the method used to estimate the
outflow's mass
(see AG for detail).
First, in order to estimate the ratio of \cooz to \thcooz as a
function of velocity,
we calculate average spectra of \thcooz and \cooz over the region
where most of the
outflow emission is found (see Figure~6).
We did this for two different regions, one for each outflow lobe.
The average spectra of these two regions are shown in 
Figures 7{\it a} and 7{\it b}.
In Figures~7{\it c} and 7{\it d}, we present the ratio  of  \cooz
to \thcooz main beam temperature ($T_{mb}^{12}/T_{mb}^{13}$), 
as a function of velocity for the average spectra of  the  
south (red) and north (blue) lobe
regions, respectively.
The line ratios, hereafter denoted $R_{12/13}(v)$, were
each fit with a second-order polynomial, as described below.

The fits to $R_{12/13}(v)$ are each a truncated parabola with fixed 
minimum at the
velocity where the average
\thcooz spectrum peaks (this velocity is obtained by a  
two-Gaussian fit
to each of the average \thco spectra in Figures~7{\it a} and 7{\it b}).
We exclude the three velocity channels closest to the  \thco peak velocity
from the fit to $R_{12/13}(v)$, as they are the velocities at which
the \thco emission might be slightly
optically thick and  where the \twco emission is probably
extremely optically thick.  In both the red and the blue lobe's 
average \thco spectra
there are ``contaminating'' velocity components
(e.g., cloud X; see Figure~7{\it a} and 7{\it b}).  
This contamination is at blueshifted
velocities in the red lobe, and redshifted velocities for the blue 
lobe.  So, after
editing for this contamination, eliminating low signal-to-noise 
points, and the three
points near the line core, we use only the points shown as filled 
symbols in Figures~7{\it c} and 7{\it d} in the final fits. 
The $R_{12/13}(v)$ parabolic fit is then used to
extrapolate $R_{12/13}$ to the high-velocity wings
of the outflow, where the \thco line is too weak to be reliably 
detected (see below).

To calculate the outflow mass at a given ($x$, $y$) position,
we directly use the \thco emission at low outflow velocities. 
 At high outflow
velocities, or wherever the \thco has not been reliably detected, we 
use the \cooz
data and the fit to $R_{12/13}(v)$, to estimate the value of
$T_{mb}^{13}(v)$ at the given velocity and
position, using the simple relation
$T_{mb}^{13}(x,y,v)=T_{mb}^{12}(x,y,v)[R_{12/13}(v)]^{-1}$.
The function $R_{12/13}(v)$ is truncated at a value of 62, the
assumed isotopic ratio
(Langer \& Penzias 1993).
Once we estimate a value of $T_{mb}^{13}(x,y,v)$, we can obtain a value of the
\thco opacity ($\tau_{13}(x,y,v)$), from which we then obtain a value
of the \thco column
density ($N_{13}(x,y,v)$) and then the mass, using Equations 1, 3,
and 4 of AG.

It should be noted that the \thcooz excitation temperature ($T_{ex}$)
is needed in order to obtain
an estimate of $\tau_{13}(x,y,v)$ from $T_{mb}^{13}(x,y,v)$ (see
Equation 1 of AG).
A value of $T_{ex}$ is obtained by assuming that the excitation
temperature for \cooz and
\thcooz are the same, and that the \cooz emission is optically thick 
at the line core.
To obtain $T_{ex}$, we use an average spectrum of the \cooz data over a region
5.75\arcmin \/ by 9.58\arcmin, centered on
20$^h$45$^m$24.7$^s$, 67\arcdeg42\arcmin19\arcsec \/ (B1950).
This  region is where most of the \thco emission lies, and thus we
are confident that
the \twco emission at ambient cloud velocities in this region is
extremely optically thick.
The peak main beam temperature of the average spectrum is about 7.1~K,
and using Equation 2 in AG, we find that $T_{ex} \sim 10.5$~K.

The value of $T_{ex} \sim 10.5$~K should be taken as a lower limit,
as other studies have shown gas in most outflows is
usually warmer (by a factor of a 1 to 10) than the ambient cloud gas
(e.g., Snell, Loren, \& Plambeck 1980;
Fukui et al.~1993; Bence, Richer, \& Padman 1996; Hatchell, Fuller \&
Ladd 1999; Davis et al.~2000).
Our high spatial resolution data taken with the IRAM~30~m telescope 
(see Paper II)
include simultaneous observations of the \coto and
\cooz lines, at the same position, which can be used in concert to
obtain an estimate of
$T_{ex}$ (e.g., Levreault 1988).
Our estimates (see Paper II for specifics) indicate that in most places
$10\lesssim \vert T_{ex} \vert \lesssim 15$,
   and that there are only very few positions with $T_{ex} > 30$~K.
Changing the excitation temperature from 10.5~K to 15~K, the outflow
mass estimate would increase by a factor of only 1.1.  A change in 
$T_{ex}$ from
10.5~K to 30~K, increases the outflow mass estimate by a factor of 1.76.
Thus,  since most of the outflow gas is at
temperatures between 10 and 15~K, assuming  $T_{ex} = 10.5$~K
does not lead to a significant uncertainty in the outflow mass estimate.

\subsubsection{Mass-Velocity Relation}

Outflows usually show a mass-velocity relation (mass spectrum)
in which \\
$dM(v)/dv~\propto~v^{-\gamma}$, and $\gamma$ may differ for 
``high" and ``low"
velocities in the flow (see Arce \& Goodman 2001a).
Using the method described above, we calculate the mass per
0.22~km~s$^{-1}$-wide
velocity channel  in the redshifted and blueshifted lobes of the
HH~315 molecular outflow.
The southern lobe redshifted gas mass as a function of outflow
velocity  is shown in
Figure~7{\it e}. The outflow velocity ($v_{out}$) is defined as the observed
velocity ($v$) minus the ``ambient" molecular cloud velocity in the
redshifted lobe region
($v_{amb, south}=2.50$~\kms, see \S 3.3 and Table~1).
We confidently detect redshifted outflow emission at outflow velocities
between  0.71 and  3.79~\kmss (that is, from the
0.22~km~s$^{-1}$-wide channel centered
at outflow velocity 0.82~\kmss to the channel centered at
$v_{out}=3.68$~\kms).
At outflow velocities lower than 0.71~\kmss the structure of the
molecular gas (both \twco and
$^{13}$CO) is nothing like the
structure of the higher velocity outflowing gas, and resembles the PV
Ceph cloud structure.
Therefore, we are confident that all of the molecular gas at
$v_{out} > 0.71$~\kmss
is outflow emission. The upper value of the redshifted velocity
($v_{out} = 3.68$~\kms)
is constrained by the sensitivity of our observations.
Figure~7{\it e} shows that the
observed mass (per 0.22~km~s$^{-1}$-wide velocity channel)
has a single power-law dependence on velocity for
the range of outflow velocities in which we detect redshifted outflow emission.
A power-law fit  yields a slope of $-3.4\pm0.2$.
We estimate a total redshifted outflow mass, within the box shown in Figure~6,
in the velocity range $0.71<v_{out}<3.79$~\kms, of
1.8~M$_{\sun}$.

The northern lobe blueshifted gas mass as a function of outflow
velocity is shown in Figure~7{\it f}.
We detect blueshifted outflow emission at outflow velocities between
-0.71 and -4.45~\kms. Similar to the southern lobe region, outflow
velocities in the
northern region are defined as the observed velocity minus the
``ambient'' central velocity.
In the northern outflow region the ambient velocity is $v_{LSR,
north}=1.5$~\kmss (see \S 3.3
and Table 1).
   At outflow velocities slower than -0.71~\kmss the structure of the \twco
starts to be much more extended than the \twco bow-shock like structure
seen at faster outflow velocities, and
thus we do not include any emission at $v_{out}$ slower than
-0.71~\kmss in our blue lobe mass estimate.
The CO data  we use in this paper show low level blueshifted outflow
emission at $\vert v_{out} \vert >
4.45$~\kms, but not at high enough signal-to-noise levels to study it.
The higher resolution data in Paper II show the high velocity structure of
the CO outflow clearly,
   and we discuss its importance there.
  Figure~7{\it e} shows that the
observed mass per 0.22~km~s$^{-1}$-wide velocity channel (or mass spectrum)
has a power-law dependence on  outflow velocity
   for $0.8\lesssim \vert v_{out} \vert \lesssim 1.5$~\kms, with a
slope of $-2.2\pm0.1$,
   and for the velocity range of  $1.5\lesssim \vert v_{out} \vert
\lesssim 4.2$~\kms,  the mass spectrum
   has slope of $-3.7\pm0.1$.
We obtain a total blueshifted outflow mass in the velocity range
$-4.45<v_{out}<-0.71$~\kms, within
the box shown in Figure~6, of 4.1~M$_{\sun}$.

The blueshifted molecular outflow lobe of HH~315
has a $\gamma = 2.2\pm 0.1$ for velocities $-1.59<v_{out}<-0.71$~\kmss
(see Figure~7{\it f}). At this velocity range the only molecular outflow
structure observed
is the bow shock-like structure coincident with HH~315C.
This structure was presumably formed,
{\it solely}, by the interaction of the HH~315C mass ejection event
with the ambient cloud.
Thus the value of $\gamma = 2.2\pm 0.1$ is consistent with
   the value predicted for an outflow created by one mass ejection episode
(see Arce \& Goodman 2001a; Matzner \& McKee 1999). 
The episodic nature of the HH~315 outflow is further
discussed in Paper II.

The \coto channel maps (Figure~2, $v_{chan}=2.98$, 3.63~\kms) show
that there is localized
redshifted emission
   in the same region of the main blueshifted outflow lobe (near HH~315C) .
We mentioned earlier (\S 3.1) that it is likely that this redshifted emission
comes from gas that has been accelerated by the ``back side'' of the
HH~315C bow shock.
Therefore, it should also be considered as outflow emission, and
should be included in the
mass and energy estimates of the outflow. Hereafter we refer to the
blueshifted and redshifted
outflow emission north of PV Ceph as the northern lobe. In Figure~6
we indicate the region used to
calculate the mass of the northern lobe redshifted emission. The
average \cooz and \thcooz spectra
in this region (see Figure~8) indicate that the ratio of these two
lines ($R_{12/13}$)
is approximately constant in the velocity
range where there is redshifted outflow emission ($2.55<v<3.65$~\kms),
with an average line  ratio ($\bar{R}_{12/13}$) of 10.
We therefore estimate the  redshifted outflow mass in the north lobe
using the same method
   described above, but instead of assuming a velocity-dependent
\cooz line opacity, we assume that the opacity is constant with
$\bar{R}_{12/13}=10$.
The resultant molecular outflow mass of the northern lobe redshifted
emission over the
outflow velocity range $1.05<v_{out}<2.15$, in the area shown in
Figure~6, is 0.7~M$_{\sun}$.

Unlike AG (where we study the HH~300 outflow),
 we do not  subtract for the ambient cloud emission
contribution to the
outflow emission at low outflow velocities (see \S 3.3.3). In
the case of the
HH~315 molecular outflow there is no indication that at the chosen outflow
velocities there is ``contamination'' from the ambient cloud
emission. As discussed above,
the molecular gas emission (at the velocities used for calculating
the outflow mass) has a structure very different from that of the
ambient cloud and a structure
which is consistent with that of an outflow lobe. Thus, we are
confident that we {\it do not}
need to correct our HH~315 CO outflow mass estimate by the ``contamination'' of
ambient cloud emission.

\section{Discussion}
\subsection{Mass and Energetics of the Outflow}
\subsubsection{The southern lobe}
The total mass of the southern (redshifted) lobe is 1.8~M$_{\sun}$.
The mass of the PV Ceph cloud (as defined in \S 3.2, also see Figure~5)
is about 74~M$_{\sun}$. Hence,
the  detectable redshifted outflow lobe mass is only $\sim$ 2\% of
the PV Ceph cloud mass.
The mass of the PV Ceph cloud in the redshifted lobe area (as defined
in Figure~6) is
29~M$_{\sun}$. Even in this localized area,  the detectable redshifted
outflowing mass is only about 4\% of the cloud mass.
The measured (line-of-sight) momentum ($\Sigma m(v_i) v_i$)
of the redshifted outflow lobe,
for $0.71<v_{out}<3.79$~\kms, is  2.1~M$_{\sun}$~\kms.
The true momentum in the redshifted outflow lobe
    should be
substantially more, as we have only considered
the line-of-sight velocity component. If we assume that the angle,
$i$, between the plane of the
sky and the outflow's ``average'' axis  is
about 10\arcdeg \/ (GKW), then the outflow momentum would
be about  12.1~M$_{\sun}$~km~s$^{-1}$.
   Our ignorance of the exact value of $i$ brings large uncertainties
to the value
of the outflow momentum. For example, a change in the value of $i$
from 10\arcdeg \/
to 15\arcdeg \/ changes the value of the red lobe momentum from
12 to 8~M$_{\sun}$~km~s$^{-1}$.
The kinetic energy is even more uncertain, as it depends on $(\sin i)^{-2}$.
We estimate the kinetic energy [$(\onehalf)~\Sigma m(v_i) v_{i}^{2}$]
   of the red lobe, for $0.71<v_{out}<3.79$~\kms, to be
$2.9~(\sin i)^{-2} \times 10^{43}$ erg. Using $i = 10$\arcdeg, the
kinetic energy
is then $9.6 \times 10^{44}$ erg.

\subsubsection{The northern lobe}
The mass
of the blueshifted gas at the outflow velocity range
$-4.45<v_{out}<-0.71$~\kmss
in the north lobe of the PV Ceph molecular outflow is 4.1~M$_{\sun}$.
If we include the 0.7~M$_{\sun}$
of  outflowing redshifted emission (at $1.05<v_{out}<2.15$~\kms)
detected near the HH~315C optical knot,
then the total mass of the northern outflow lobe is 4.8~M$_{\sun}$.
We estimate  the mass of the Northern Cloud (see \S 3.3),
from the \thco emission, to be 13.5~M$_{\sun}$.
Comparing this amount of mass, with the total northern lobe mass, we
see that the outflowing mass in the north
region is about one third of what one would naively consider the
``ambient'' cloud mass.
Later, we show that about half of the mass of the medium-density gas
which we at first naively define as the
ambient north cloud gas, is in fact gas that has been accelerated by
the HH~315 flow (see \S 4.2.2).

We obtain a line-of-sight momentum of 5.1~M$_{\sun}$~\kmss for the blueshifted
outflow emission and a line-of-sight momentum of 1.0~M$_{\sun}$~\kmss for the
redshifted outflow emission in the north lobe, for a sum of
6.1~M$_{\sun}$~\kms.
Assuming $i=10\arcdeg$, then we obtain
a total north lobe momentum of 35.1~M$_{\sun}$~\kmss. The kinetic
energy estimates for the
north lobe are $7.6~(\sin i)^{-2} \times 10^{43}$ erg for the
blueshifted gas and
$1.5~(\sin i)^{-2} \times 10^{43}$ erg for the redshifted gas. If we
assume $i=10\arcdeg$,
then the total
kinetic energy of the north lobe is $3.0 \times 10^{45}$ erg.
The mass, momentum, and kinetic energy
estimates of the HH~315 outflow are listed in Table~2.

\subsection{The effects of the outflow on the ambient gas}

\subsubsection{The southern lobe and the PV Ceph Cloud}
As shown in Figure 5, the projected overlap between outflowing gas 
and the  PV
Ceph cloud (as defined in Table 1) is quite small.  Nonetheless, it is still
interesting to consider what effect(s) the outflow may have on the 
overall evolution
of the PV Ceph cloud.  Since the large-scale
\twco velocity maps in this paper (Figure~2) show no blueshifted
outflow lobe emission superimposed on the PV Ceph cloud we  only 
consider the
effects of the redshifted lobe in estimating the effects of the 
HH~315 flow on the
dense portion of the star-forming cloud. Note, though, that our
 higher resolution observations (Paper II)  {\it do}  detect  a 
small amount
of blueshifted outflow emission just north of PV Ceph.

One method to quantitatively study the effects of an outflow on its
parent cloud is to
compare the outflow's  energy  with the cloud's binding energy.
The cloud's binding energy is given by
$E_{grav} \sim GM_{c}^{2}/R_c$, where we estimate
$R_c$ to be the geometric mean of the short and long axes of the PV Ceph cloud
($\sim 0.7 $~pc), and $M_{c}$, the mass of the cloud, is $\sim
74$~M$_{\sun}$.
Using these values we obtain that the cloud's binding energy
is about $6.7\times 10^{44}$~erg.
If the HH~315 flow has an inclination to the plane of the sky of  $i\sim
10\arcdeg$,  then the southern lobe of HH~315 has enough kinetic energy
($9.6\times 10^{44}$~erg) to gravitationally
unbind its parent cloud ---if that energy is efficiently coupled to
the cloud (see below).
Given the substantial amount of kinetic energy in the southern lobe,
the flow might be able
to disperse a major fraction of the very dense gas, where most of the
binding mass is located.
This in turn would alter the gravitational potential well of the cloud.

We may indeed already be seeing the start of the PV Ceph cloud's
disruption by the HH~315 flow.
In Figure~9{\it a} we plot, in contours, the integrated intensity of the
HH~315 molecular outflow's south lobe,
superimposed on the \thco integrated intensity over the
velocity range $1.84<v<2.17$~\kms. It can be seen that the molecular
outflow's southern lobe fits in
the \thco cavity seen just south of PV Ceph. This spatial coincidence
suggests that the cavity was formed
by the the outflow-cloud interaction. In addition,
the peak \thco integrated intensity over the velocity range $2.17<v<2.61$~\kmss
   borders the southern lobe's position (see Figure~9{\it b}). This structure
in the \thco also suggests that the southern
lobe has pushed and redistributed the medium-density ambient gas.

In order to further investigate the hypothesis that the southern
molecular outflow
   lobe is responsible for the minimum
in the \thco distribution observed south of PV Ceph, we studied
the velocity distribution of the \thco in the region
(see spectrum in Figure~10).
We made an
average of the \thco emission in the cavity and  fit it with a
two-Gaussian function (the small
blueshifted velocity component is due to cloud X). The fit yields a
central velocity of 2.6~\kmss for the
velocity component associated with the PV Ceph cloud. This is only
0.1~\kmss larger than
   $v_{amb, south}$. Even though the difference is small, it is still
significant as
the 1-$\sigma$ error in the central velocity fit is about 0.01~\kms.

In Figure~11 we show (in white contours) a \coto position-velocity ($p-v$)
diagram superimposed on a  grey-scale
\thcooz $p-v$ diagram. Both $p-v$ diagrams are constructed by 
summing all spectra over the width of the \thco cavity seen in
Figure~9{\it a}.
 The cavity is seen as a minimum
in the \thco emission 
 at velocities between $\sim 1.5$ and $\sim 2.2$~\kms, at
0 to 2.5\arcmin \/ offset from the position of PV Ceph.
This minimum in the
\thco emission (or cavity) is bordered to the north and south by emission
from the  PV Ceph cloud. The redshifted ``border'' of the cavity is 
formed by the emission
of the PV Ceph cloud and the blueshifted border is defined by emission from 
Cloud X. At the same position  of the cavity there is a notch in the \coto
$p-v$ diagram. This notch indicates that there is also a minimum in the \coto
emission at the position and velocity of the \thco cavity, 
and that there is a bulk shift
in the  velocity distribution of the \coto gas towards  redshifted velocities.

One possible scenario which fits the data is that the south molecular
(redshifted) lobe pushed and
accelerated the ambient cloud gas. The cavity can only be seen at low
blueshifted velocities
(relative to the PV Ceph cloud's central velocity, $v_{LSR,
south}=2.50$~\kms)  since
the redshifted outflow lobe has only been able to  slightly change
the velocity distribution (toward redshifted velocities)
of the dense gas it has interacted with, and it yet has to clear the
region completely.
If the outflow does not turn off, then given more time, more ambient
gas will be shoved away and
   accelerated, making the cavity larger and visible at all velocities.

We may also value the role of the HH~315 flow in the PV Ceph cloud
by comparing the southern lobe's kinetic energy with the cloud's
turbulent kinetic energy.
The total kinetic energy of a cloud where thermal motions are a negligible
part of $\Delta V$ (the observed FWHM velocity line width),
is given by $E_{turb}=\frac{3}{16\ln 2} M_{c} \Delta V^2$, where
$M_{c}$ is the cloud mass.
   The average \thco velocity width in the PV Ceph cloud  is
$\Delta V \sim 0.6$~km~s$^{-1}$, so $E_{turb} \sim 1.2 \times 10^{44}$~erg.
The outflow kinetic energy is about $2.9~(\sin i)^{-2} \times
10^{43}$~erg  in the southern lobe,
   thus, even if
$i=29\arcdeg$, the southern lobe kinetic energy exceeds the cloud's
turbulent kinetic energy. Similar to what is observed in the giant
molecular outflow HH~300 (AG), 
the southern lobe of the HH~315 molecular outflow has enough
kinetic energy to
potentially feed the turbulence in its parent cloud.

Another way to estimate the impact of the HH~315 outflow on the PV Ceph cloud
is by comparing the outflow's energy with the chemical binding energy
of the molecules in the cloud. If we assume $i=10\arcdeg$, then the kinetic
energy of the molecular outflow is about $9.6 \times 10^{44}$~erg. This
energy would be enough to dissociate about $10^{53}$
 H$_2$ molecules (each with a molecular binding energy of 4.5 eV).
This amount of molecules is equivalent to 0.2~M$_{\sun}$, 
which is only a mere 0.3\% of the 
total mass of the cloud. Therefore, the potential disruption of
molecules by the HH~315 flow will have  no significant impact on the PV Ceph
cloud as a whole. 

Yet another way to asses the importance of the HH~315 outflow
energy input on its parent cloud is by comparing the flow's power
with the power
needed to maintain the magnetohydrodynamic (MHD) turbulence in the 
PV Ceph cloud.
   The three-dimensional numerical simulations of Stone, Ostriker, 
\& Gammie (1998) 
and Mac Low (1999) study
MHD turbulence under density, temperature and magnetic field conditions
representative of those found in Galactic molecular clouds.
 Using Equation 7
in Mac Low (1999),
we can estimate the input power needed to
 maintain MHD turbulence in the PV Ceph cloud
(shown in Figure~5).
We use the estimated mass of the PV Ceph cloud (74 M$_{\sun}$) and 
the average velocity line width of the \thco (0.6~\kms).
Using these values we find that about  0.01 L$_{\sun}$ 
of input power at a scale
of 0.5~pc (the approximate length of the southern lobe) are needed
to counter the dissipation of MHD turbulence in the
PV Ceph cloud.

The power of an outflow is usually estimated by dividing the
outflow kinetic energy by the dynamical age of the outflow.
The conventional way to estimate the dynamical age of a molecular
outflow assumes
that all the gas in the outflow originates at the young star. This
assumption is wrong, since the vast majority of the gas in a molecular outflow
comes from the entrained gas in the host cloud (along the extent of
the outflow)
   that has been put in motion by
the underlying stellar wind. Since there is no way to obtain
an accurate estimate of the outflow lifetime, we
estimate a lower and an upper bound to the dynamical age in order to estimate
an upper and lower bound to the outflow power.
We estimate the lower limit to be
the time it has taken HH~315F (the HH knot which currently lies farthest from
the source) to travel to its current position. The distance from PV Ceph
to HH~315F is 1.41~pc (RBD), and if
we assume a tangential velocity of 200~km~s$^{-1}$,
   we then obtain a dynamical age of about 6900~yr. As an upper limit
on the age we use
$2 \times 10^{5}$ years, the statistical lifetime of outflows as derived
by the study of Parker, Padman, \& Scott~(1991).
We then estimate the lower and upper limits of the outflow power to be
$1.2 \times 10^{-3}~(\sin i)^{-2}$ to
$3.4 \times 10^{-2}~(\sin i)^{-2}$~L$_{\sun}$.
If we assume that $i\sim 10\arcdeg$,
then the southern lobe's power is more than the estimated power needed
to maintain the MHD turbulence in the PV Ceph cloud.

Note that we only  compare the outflow power to that needed to
sustain the MHD turbulence
and stop gravitational collapse in the cloud, to quantify the
possible effect that outflows have on their
parent cloud: by no means we are implying that molecular clouds {\it
need} to be maintained in gravitational
equilibrium.  Our results for HH~315 (this paper) 
and HH~300 (AG) imply
that outflows are sources of non-negligible
power in clouds, and they should be treated as such in numerical
simulations of MHD turbulence in star-forming
clouds.

The energy in the HH~315 outflow southern
lobe is comparable or larger than the
binding energy, and the turbulent energy, of the PV Ceph cloud.
But, this does not necessarily mean that the outflow will be able to
unbind the cloud or
turn all its energy into cloud turbulent energy. We would be able to
make a rough prediction of the
PV Ceph cloud's future from our observations, if we knew how well the
outflow energy couples
to the cloud.
The \thco cavity coincident with the southern outflow lobe,
   the \thco shell structure in the north, coincident
with the outer edge of the northern blueshifted lobe, and
the velocity gradient in the  velocity of the ambient gas
   (see below) are all examples that there is some
degree of coupling between the outflow and the ambient gas.
Just from observational data, though,  it is very hard to predict the
outflow-cloud coupling efficiency
of any outflow. No sufficiently good episodic outflow-cloud
interaction model exists (see Arce \& Goodman 2001a).
There is a need for
theoretical and numerical studies which concentrate on studying how
efficiently can bulk
motions, produced by outflows, unbind a cloud and/or  produce turbulence,
as a function of cloud density and outflow energy.

\subsubsection{The northern lobe and the Northern Cloud}

The northern lobe of the PV Ceph  molecular outflow is characterized
by its bow shock-like structure,
coincident with the optical knot HH~315C, most evident at low outflow
velocities
(see Figure~2). A shell-like structure, surrounding the blueshifted outflow lobe,
can be seen in \thco emission at very low (between 0.79 and
1.45~\kms)  velocities
(see Figures~3{\it a} and 12).
Most of the northern lobe
region is essentially devoid of \thco emission, and the majority of
the \thco emission present in the
region surrounds the western and northern edge of the \twco bow shock
structure (see Figure~12).
It appears that
the HH~315 flow has been able to push gas 
north of PV Ceph, piling it in a dense  
 shell-like
structure surrounding the outflow lobe, which we detect in $^{13}$CO(1--0).

Using the \thco emission surrounding the blueshifted lobe at velocities between
0.79 and 1.45~\kms,
we estimate the shell structure's  mass to be
6.8~M$_{\sun}$ and momentum to be  2.5~M$_{\sun}$~\kms.  The 
line-of-sight momentum
of  the shell is sightly smaller
than the line-of-sight momentum of the blueshifted outflow emission
   (which is equal to  5.1~M$_{\sun}$~\kms).
The morphology of the shell-like structure (which surrounds the
blueshifted outflow lobe), its slow
blueshifted velocity, and the fact that its momentum is at least similar
to the blueshifted outflow
momentum are all consistent with the \thco shell having been formed
by the HH~315's
(momentum-conserving)
entrainment of the ambient molecular gas.

The velocities at which the \thco shell structure is seen fall inside
the range of velocities
which we defined earlier
(\S\S 3.3 and 4.1.2) as the north cloud ambient cloud. Thus,
in fact we see that a substantial fraction ($6.8/13.5 \sim 50$\%)
of what we had earlier naively defined as ``Northern Cloud ambient gas'' in
   reality is medium density gas that has been accelerated
by the outflow. In fact, the peak \thco emission shell structure can
be nicely seen
   in the \thco $p-v$ diagram (Figure~4),
at about 6\arcmin \/ ($\sim 0.75$~pc) from the source (along the
P.A.=-26\arcdeg \/ axis),
   at a velocity of about 1.35~\kms.
Thus, most of the cloud structure in the northern lobe region, as
defined by the \thco emission, comes
from gas that has been greatly affected by the outflow. One might say
that in this region
   ``the cloud is the outflow and the outflow is the cloud''.

   In Figure~13 we plot the
   \coto  position-velocity ($p-v$) diagram of the mapped region,
constructed by summing all spectra along an axis with a position angle
of -26\arcdeg. By summing the \coto spectra over a limited width (see
Figure~6) of the \coto map, we avoided including
most (but not all) of the region where cloud X lies, 
so that its emission would not
``contaminate'' the \coto $p-v$ diagram.
   The main features of the \coto $p-v$
diagram are the peaks in the maximum velocity associated with the
blueshifted (north) and redshifted
(south) lobes. A  local increase in the maximum velocity may be seen
at the position (along the
average axis) of the optical HH knots A, B, C, and D, and close to
the position of PV Ceph.
The velocity structure at the position of each of these optical knots is
characteristic of the bow shock
(prompt) entrainment mechanism, which will produce
the highest velocities at the head of the shock (the position of the HH knot)
and decreasing velocity trend towards the source (e.g., Bence et
al.~1996; Lee et al.~2000; 2001).
The fact that we see several of these velocity increases (or bumps)
in the $p-v$ diagram, and each is
coincident with a different HH knot is also evidence that HH~315 flow
is an episodic outflow (see Arce \& Goodman 2001a).
The increase in redshifted velocities at about -1\arcmin \/
offset from the source position
is mainly due to the redshifted outflow lobe.
We further discuss this velocity feature, and the
kinematics of HH~315 molecular outflow  in Paper II.

In addition to the kinematical structure associated with the HH
knots, the north half of the $p-v$ diagram shows
evidence for a velocity gradient in the ambient gas' central velocity
($v_{LSR}$). The ambient cloud's $v_{LSR}$
may be approximated as  the velocity where the peak contour, in the
$p-v$ diagram, lies. In the \coto $p-v$
diagram, the peak contour is seen to shift from about 2.5~\kmss at 
the position of PV
Ceph down to  about 1.5~\kmss near HH~315A.
We believe that the northern lobe of the HH~315 outflow is responsible for the
gradient in the central velocity of the ambient cloud.
In addition, the fact that more than half
of the gas mass in the northern region has been accelerated by the
outflow gives additional
support to our hypothesis.

In any volume of momentum-conserving wind-cloud interaction 
high-density gas will be
accelerated less than gas at low densities. This is what we observe in the
ambient gas surrounding HH~315.
The ambient gas north of PV Ceph (where we detect an obvious velocity gradient)
is much less dense than the ambient gas to the south (where we cannot 
easily detect a
large-scale velocity gradient in the CO emission).
The ambient density contrast between north and south 
of PV Ceph is clear from our
\thco integrated intensity map (Figure~5) and from the 
relative extinction measurements between
the region around HH~315B and HH~315D (GKW). We may also estimate
the density contrast, from the ambient gas
velocity gradient 
 in each region. We assume that the
 velocity gradient in the ambient gas central velocity
 ($\Delta v_a$)
 is wholly 
due to the momentum conserving stellar wind-ambient gas interaction, 
and that the stellar wind 
deposits all its momentum ($\rho_w v_w$) on the ambient gas, so that 
$\rho_w v_w \approx \rho_a \Delta v_a$. We assume that the wind 
momentum of each lobe 
is the same. The \coto shows a velocity gradient to the north 
of $\sim 1$~\kmss (see Figure~13).
The \coto $p-v$ diagram shows no velocity gradient in the ambient gas
south of PV Ceph
 due to the coarse velocity resolution 
(0.65~\kms) of the \coto data.
 However,  the \thco data shows a velocity shift 
of $\sim 0.1$~\kmss in the southern molecular outflow lobe region
(see \S 4.2.1, and Figure~10).
Thus, we estimate that the 
 ambient gas density 
south of PV Ceph is about a factor of 10 larger than the 
 ambient gas density  north of PV Ceph.

\section{Summary and Conclusion}

We mapped the giant molecular outflow
associated with the HH~315 flow, from the star PV Cephei, in the
\coto line, with a
beam size of 27\arcsec. We also made  more extended maps
of the gas surrounding the HH~315 flow in the \cooz and
\thcooz lines, with  45\arcsec \/ and 47\arcsec \/ beam sizes, respectively.
By observing a large extent of the gas surrounding the
outflow we are able to study the outflow in the context of
its surrounding medium. Also, the \thco observations help us
assess the effects the outflow has on the surrounding
moderate-density ($n \sim 10^3$~cm$^{-3}$) gas structure and kinematics.
In addition, the combined \cooz and \thcooz line observations
enable us to estimate the mass of the outflow
by correcting for the velocity-dependent opacity
of the \cooz line.

The outflow source, PV Ceph, is a young star forming in the
northern {\it edge} of
its dark cloud. Hence, the northern
and southern lobes of the HH~315 outflow are interacting
with environments with drastically different densities.
The southern lobe has a mass of 1.8~M$_{\sun}$ and
interacts with the dense region we define as the PV Ceph cloud.
We quantify the effects that the HH~315 flow has on
the PV Ceph cloud by using different methods.
Assuming that HH~315 has an inclination to the plane of the sky ($i$)
of about 10\arcdeg, then the southern lobe of the HH~315 molecular
outflow has a momentum of  12.1~M$_{\sun}$~\kmss
and a kinetic energy of $9.6 \times 10^{44}$~erg.
The kinetic energy of the southern outflow lobe is enough  to
supply the turbulent energy of the PV Ceph cloud
($1.2 \times 10^{44}$~erg), and
enough energy to gravitationally unbind it (the gravitational binding energy
of the PV Ceph cloud is $\sim 6.7 \times 10^{44}$~erg).
We compare our estimate of the southern lobe's mechanical power, with that
of the power needed to sustain the MHD turbulence in the PV Ceph cloud
(using the results of the
numerical models of Stone, Ostriker, \& Gammie~1998 and Mac Low~1999).
If we assume that $i\sim 10\arcdeg$,
then the southern lobe's power is more than the estimated power needed
to maintain the MHD turbulence in the PV Ceph cloud.
In addition, the $^{13}$CO data indicate that the southern HH~315
flow is having
a major effect on the distribution of the ambient cloud gas. We detect
a cavity in the distribution of the \thco in the PV Ceph cloud region. The
\thco gas velocity and density distribution, and the morphology of
the outflowing \coto gas
suggests that the cavity has been formed by the southern outflow lobe
interaction with its surroundings.

The northern (mostly blueshifted) lobe has a total mass of 4.8~M$_{\sun}$ and
interacts with an environment which is much less dense (by about a
factor of ten)
than the environment south of PV Ceph.
There is much less \thco emission north of PV Ceph, and most of the
\thco emission is concentrated
in a shell-like structure about 1.2~pc northwest of PV Ceph,
coincident with the
outer edge of northern outflow lobe. It appears that the northern
lobe of the HH~315 molecular outflow
has ``pushed'' aside the $^{13}$CO, piling the gas
in a shell-like structure at its edges.
The morphology of the shell-like structure, its slow
blueshifted velocity, and the fact that its momentum is very similar
to the blueshifted outflow lobe
momentum are all consistent with the shell being formed by the
(momentum-conserving)
entrainment of the ambient molecular gas in
the HH~315 flow.
Adding the mass in the northern molecular outflow, and the mass of
the \thco shell, we find that more than
50\% of the total gas mass in the region of the northern outflow lobe
has been put into motion by the outflow.
The northern outflow lobe of HH~315 is having a major impact on the
structure of the ambient medium, even at a distance
of about 1 to 1.4~pc from the outflow source.

In addition, we detect a velocity gradient in the ambient cloud's
central velocity north of PV Ceph.
The gradient is towards
blueshifted velocities, in the same sense as the vast majority
   of the outflowing gas in the northern outflow lobe.
We argue that the gradient in the cloud's central velocity is due to
the HH~315 outflow, as the gradient is
consistent with it being formed by a momentum-conserving
outflow-cloud interaction. The fact that the outflowing
gas mass north of PV Ceph is a major fraction of the total gas mass,
supports our picture in which the HH~315
flow is responsible for the velocity gradient in the ambient gas
north of PV Ceph.

Our study clearly shows that  the HH~315 flow is effecting the
kinematics of its surrounding medium, and has been able to redistribute
considerable amounts of its surrounding medium-density
($\sim 10^3$~cm$^{-3}$) gas in its star-forming core
as well at parsec-scale distances from the source.
These are necessary steps in the clearing of the gas surrounding a forming star
and the eventual (total) disruption of the parent cloud.
We conclude that the giant outflow from PV Ceph is having a
profound effect on its host cloud's evolution and fate.
It is tempting to generalize this result, along others like it, and
suggest that outflows have
great influence on the sculpting of star-forming molecular clouds.

\acknowledgements
We would like thank Paul Ho, and 
Charlie Lada for their helpful comments on this work.
We would also like to thank  Bo Reipurth for his comments and the optical 
image of PV Ceph.
And we are grateful to the National Science Foundation 
for supporting
this effort through grants AST 94-57456 and AST 97-21455.

\clearpage

\figcaption[figure1.eps]{Integrated intensity contour map of the
\coto HH~315 outflow superimposed on  a
wide-field H$\alpha$+[S~II] (optical) CCD image from RBD. The
blueshifted lobe, which is integrated over the
velocity range $-2.55<v<0.7$~\kms, is represented by the light grey
contours. The first contour and contour steps of the blueshifted lobe
are 1.04 and 0.52~K~\kms, respectively.
We cut out the contribution from cloud X (see text) from the integrated
intensity map of the blueshifted lobe, hence the
cut contours at the
westernmost edge of the lobe. The redshifted emission is represented
by the dark contours.
The redshifted emission in
the north lobe is integrated over the velocity range $2.65<v<3.30$~\kms.
The southern redshifted lobe is integrated over
the velocity range $3.30<v<6.55$~\kms. The first contour and contour
steps of both
north and south redshifted integrated intensity emission all have a
value of 0.52~K~\kms.
The position of the HH knots in the optical map, and the
position of PV Ceph are shown. The extent of the OTF map is indicated
by the jagged thin line.
The FWHM beam of
the NRAO 12~m telescope is shown on the bottom-right corner. \label{intco21.hh315}}

\figcaption[figure2.eps]{Velocity channel maps of the \coto emission.
Each velocity channel is 0.65~\kmss
wide. The center velocity of the channel is shown on the upper-left
corner of each panel.
The starting contour and the
contour steps are both 0.325~K~\kmss for all panels.
The position of the outflow source (PV Ceph)
is represented by the star symbol, at the center of each panel. The
crosses show the position of the
brightest (optical) point of the different HH knots which make the
HH~315 flow, knots C, B, A, D, E, F (from top to
bottom) (RBD). We also identify the cloud structure we define as cloud
X. The linear scale, assuming a distance to the
source of 500~pc, is shown. The FWHM beam of
the NRAO 12~m telescope is also shown.
\label{co21vel.hh315}}

\figcaption[figure3.eps]{Velocity-range-integrated
   intensity maps of the \thcooz emission. The
velocity interval of integration is given on the top of each panel. The
first contour, which is also the value of the contour steps, is given
inside brackets on the bottom-right
corner of each panel in units of K~km~s$^{-1}$.
The position of the outflow source, PV Ceph, is identified
with a star symbol. The cross symbols indicate the position of the HH knots,
the same as in Figure~1.
In panel {\it (a)} we identify the filamentary structure associated
with cloud X.
We also indicate the shell-like structure
formed by the blueshifted lobe of the HH~315 outflow (see \S 4.2.2).
In panel {\it (c)}
   we show the cavity in the \thco emission
presumably formed by the redshifted (southern) lobe of HH~315 (see \S 4.2.1).
\label{13covel.hh315}}

\figcaption[figure4.eps]{Position-Velocity ($p-v$) diagram of the
\thcooz emission,
constructed by summing all spectra along an axis with a
P.A.=-26\arcdeg \/ (the outflow ``average'' axis), over
the width of the map. The \thco $p-v$ digram is obtained from a smoothed
cube of our \thco data with a velocity
resolution of 0.11~\kms, and 11.5\arcsec \/ by 11.5\arcsec \/ pixels
(see \S 2.2).
The contours are from 4 to 30~K, in steps of 2~K, and from 35 to 90~K in steps of 5~K.
The dashed square on the bottom denotes the limits of our definition
of the PV Ceph cloud (see \S 3.3).
The dashed square on the top denotes the limits of what we define in
the text as the Northern Cloud (see \S 3.3).
The vertical lines at the top and bottom denote the
velocity range of integration
of the \thco velocity maps shown in Figure~3. The letter at the
center of each velocity range
   indicates the panel in Figure~3 which corresponds to the given range.
\label{13copv.hh315}}

\figcaption[figure5.eps]{Integrated intensity grey-scale map of the
\thcooz PV Ceph cloud.
The velocity range of integration is $1.8<v<3.2$~\kms. 
The dashed square denotes our definition of the PV Ceph cloud area
(see \S 3.3).
The white contours represent the \coto
integrated intensity of the redshifted outflow
lobe. The contours are the same as in Figure~1,
except that here, we only show the 0.52 $\times$ (1, 2, 4, 6, 8,
10)~K~\kmss contours.
 \label{13coint.hh315}}

\figcaption[figure6.eps]{Map showing different regions defined in the text.
The contours are the same as the contours in Figure~1.
The dashed square around the blueshifted (northern) lobe
and around the southern (redshifted) lobe indicate the region used to
obtain the average \cooz and \thcooz spectra
shown in Figure~7. The solid square  around the blueshifted  lobe and
around the southern lobe indicate the region
used to obtain the outflow mass of each lobe. The solid square around the
redshifted emission in the {\it northern} lobe
indicates the region used to obtain the average
\cooz and \thcooz spectra shown in Figure~8, and the mass of the northern
redshifted outflow emission. The solid grey tilted rectangle shows the area
used to obtain the \coto position-velocity diagram of Figure~13.
\label{regions.hh315}}

\figcaption[figure7.eps]{
({\it a}) Average spectra over the
redshifted (southern) lobe
region indicated in Figure~6.  Dark line indicates the \thcooz
average spectrum, the light line
indicates the \cooz average spectrum. ({\it b}) The same as [{\it a}],
but for the blueshifted (northern) lobe.
({\it c}) Main beam
temperature (or intensity) ratio of \cooz to $^{13}$CO(1--0),
using the average spectra in the above panel, as a function of
observed velocity ($v$).
This ratio is denoted in the text as $R_{12/13}$.
   The filled circle symbols are the points
used for the second-order polynomial fit to
$R_{12/13}$. The solid line is the resulting fit.
({\it d}) The same as [{\it c}], but for the blueshifted lobe.
({\it e}) Mass spectrum, or mass in a 0.22~km~s$^{-1}$-wide channel
as a function of outflow velocity ($v_{out}=v-v_{amb}$),
   for the red (southern) lobe of the HH~315 molecular
outflow. The filled diamonds represent the outflow, and the unfilled
diamonds represent the
ambient cloud.
The solid line shows a power-law fit to the filled diamonds.
The slope of the fit is $-3.4\pm0.2$.
({\it f}) Similar to [{\it e}], but for the northern lobe. The dark
diamond symbols
represent the blueshifted emission. The grey squares represent the
redshifted emission in
the northern lobe.
We made two power-law fits to the blueshifted emission in the northern lobe.
   The slopes of the fits are $-2.2 \pm 0.1$ for the low-outflow velocity range
and $-3.7\pm 0.1$ for the high-outflow velocity range.
The grey line shows a power-law fit to the northern lobe redshifted
outflow emission.
The slope to this fit is $3.1 \pm 0.8$.
\label{opcor.hh315}}

\figcaption[figure8.eps]{ 
({\it a}) Average spectra over the northern
redshifted outflow emission
region indicated in Figure 6.  Dark line indicates the \thcooz
average spectrum, the light line
indicates the \cooz average spectrum. ({\it b}) Main beam
temperature (or intensity) ratio of \cooz to $^{13}$CO(1--0)
---also denoted in the text as $R_{12/13}$--- as a function of
observed velocity ($v$), using the average spectra in [{\it a}].
\label{opcornw.hh315}}

\figcaption[figure9.eps]{
({\it a}) Same grey scale map as in
Figure~3{\it c}, but with different grey scale range.
   In addition, we superimpose the
integrated intensity contours (white lines) of the \coto southern
lobe shown in Figure~1.
    Notice that the integrated intensity contours of the southern lobe
   fit well in the \thco ``cavity" (or minimum).
({\it b}) Grey scale map similar to Figure~3{\it d}.
In addition, we superimpose the
integrated intensity contours (white lines) of the \coto southern
lobe shown in Figure~1.
   Notice that the maximum \thco emission
borders the western edge of the molecular outflow lobe. 
\label{southcav.hh315}}

\figcaption[figure10.eps]{Average \thcooz spectrum over the \thco cavity 
south of PV Ceph.
The area is indicated by a dashed box in panel Fig.~9{\it b}].}

\figcaption[figure11.eps]{\coto position-velocity diagram (white contours)
superimposed on a \thco grey-scale position-velocity diagram. 
Both $p-v$ diagrams are constructed by summing all spectra
 over the width of the \thco cavity seen in
Figure 9a. 
The contours in the \coto $p-v$ diagram are 4 to 30~K, in steps of 2~K.
The contours in the \thcooz $p-v$ diagram are 
2, 3, 5, 7, 9, 11, 13, 15, 17, 19, 21~K.
The box encloses the region of the \thco $p-v$ diagram shown in more
detail in the bottom panel.
Several velocity features are identified.
 \label{cavpv.hh315}}

\figcaption[figure12.eps]{Grey scale map similar to Figure~3{\it a}. In
addition we superimpose the \coto integrated
intensity contours (white lines) of the blueshifted (northern) 
outflow lobe.
Notice the shell-like
structure in the \thco surrounding the outer edge of the \coto
outflow lobe. The \thcooz spectrum shown is an
   average over
the area inside the white dashed box. The spectrum axes have a velocity
range of $-5<v<8$~\kmss and a main
beam temperature range of $-0.1<T_{mb}<1.0$~K.  The vertical dashed
line in the spectrum
indicates the position of $v=1.5$~\kmss in the velocity axis.
\label{shell.hh315}}

\figcaption[figure13.eps]{\coto position-Velocity ($p-v$)
diagram of the HH~315 molecular outflow,
constructed by summing all spectra along an axis with a
P.A.=-26\arcdeg \/ (the outflow ``average'' axis), over
an area shown in Figure~6.
The contours are 
3 to 10~K in steps of 1~K, 12 to 20~K in 
steps of 2~K, and 25 to 85~K in steps of 5~K.
The positions, along the average outflow axis,
 of the brightest optical emission of the six major HH
knots in the flow
are marked with a horizontal solid line.
The thick black dash line traces the maximum contour along the outflow axis,
indicating the presumed ambient cloud gas 
central velocity. Notice the velocity gradient north of PV Ceph.
\label{co21pv.hh315}}

\clearpage

\begin{deluxetable}{lp{7.2cm}p{5.1cm}p{3.0cm}}
\rotate
\tabletypesize{\footnotesize} 
\tablewidth{0pt}
\tablecaption{Description of Features in the Region Surrounding HH~315 
\label{hh315features}}
\tablecolumns{4}
\tablehead{
\colhead{Feature} & 
\colhead{Description} & 
\colhead{Number Value or Range} & 
\colhead{See also}
}
\startdata
PV Ceph Cloud  & 
Medium-density gas condensation associated with the formation of PV Ceph  &
$1.8<v<3.2$~km~s$^{-1}$ 
   \newline $20^h44^m32^s<\alpha_{1950}<20^h46^m30^s$
   \newline $67^o36'30''<\delta_{1950}<67^o48'40''$ \newline &   
Figs.~4 and 5, \S 3.3 \\

North Cloud  & 
Medium-density gas condensation north of PV Ceph, associated with 
    gas of northern outflow lobe & 
$0.8<v<2.2$~km~sec$^{-1}$
  \newline area same as northern 
  \newline outflow lobe (see below) 
   \newline &
Figs.~5 and 6, \S 3.3\\
$v_{amb, south}$ & Central velocity of ambient PV Ceph cloud gas \newline &
2.5~km~s$^{-1}$ & Fig.~7{\it a}, \S 3.3\\
$v_{amb, north}$ & Central velocity of north cloud  \newline &
1.5~km~s$^{-1}$ & Fig.~7{\it b}, \S 3.3\\

Southern lobe area & 
Area used to estimate the mass of the southern molecular
  outflow lobe \newline &
$20^h45^m10^s<\alpha_{1950}<20^h46^m05^s$
  \newline $67^o41'30''<\delta_{1950}<67^o48'05''$ & 
Fig.~6, \S 3.4\\

Northern (blue) lobe area & 
Area used to estimate the mass of the 
  blueshifted molecular gas of the northern lobe \newline &
Non-square area, see Fig.~6 & 
Fig.~6, \S 3.4\\

Northern (red) lobe area & 
  Area used to estimate the mass of the redshifted molecular 
  gas of the northern lobe  &
$20^h44^m08^s<\alpha_{1950}<20^h44^m44^s$
  \newline $67^o51'30''<\delta_{1950}<67^o54'50''$ & 
Fig.~6, \S 3.4
\enddata
\end{deluxetable}

\clearpage

\begin{deluxetable}{lccccccc}
\rotate
\tablecolumns{8}
\tabletypesize{\footnotesize} 
\tablewidth{0pt}
\tablecaption{Mass, Momentum, and Kinetic Energy of the HH~315
Molecular Outflow
\label{ener.hh315}}
\tablehead{
\colhead{} &
\multicolumn{2}{c}{Southern Lobe} &
\colhead{} &
\multicolumn{4}{c}{Northern Lobe}
\\  \cline{2-3} \cline{5-8}
\colhead{} &
\colhead{Redshifted gas\tablenotemark{a}}  & 
\colhead{If $i=10\arcdeg$\tablenotemark{b} \hspace{0.5cm}} &
\colhead{} &
\colhead{Blueshifted gas\tablenotemark{c}} & 
\colhead{Redshifted gas\tablenotemark{d}} & 
\colhead{Total} & 
\colhead{If $i=10\arcdeg$\tablenotemark{b}}
}
\startdata

Mass & 1.8 & \nodata \hspace{0.5cm}& \/ & 
4.1 & 0.7 & 4.8~M$_{\sun}$ & \nodata \\
Momentum & $2.1~(\sin i)^{-1}$ & 12.1\hspace{0.5cm} & \/ &
$5.1~(\sin i)^{-1}$ & $1~(\sin i)^{-1}$
& $6.1~(\sin i)^{-1}$ & 35.1~M$_{\sun}$~km~s$^{-1}$\\
Kinetic Energy & $2.9~(\sin i)^{-2}$ & 96\hspace{0.5cm} & \/  &
     $7.6~(\sin i)^{-2}$ & $1.5~(\sin i)^{-2}$ & $9.1~(\sin i)^{-2}$ &
$302 \times 10^{43}$~erg\enddata
\tablenotetext{a}{We only detect redshifted gas in the southern lobe.
The mass, momentum,
and kinetic energy are obtained from the gas in an area shown in
Figure~6, over the velocity range
$3.21<v<6.29$~\kms.}
\tablenotetext{b}{Value of physical parameter assuming that the angle
between the plane of the
sky and the outflow ``axis'' ($i$) is 10\arcdeg.}
\tablenotetext{c}{Obtained from the blueshifted gas in the area shown
in Figure~6, over the velocity range
$-3.95<v<0.79$~\kms.}
\tablenotetext{d}{Obtained from the redshifted gas in the area shown
in Figure~6, over the velocity range
$2.55<v<3.65$~\kms.}
\end{deluxetable}

\end{document}